# Thulium-doped silica fibers with enhanced $^3H_4$ level lifetime: modelling the devices for 800-820 nm band

Pavel Peterka[*a], Ivan Kasik[a], Anirban Dhar[a,c], Bernard Dussardier[b], and Wilfried Blanc[b]

[a]Institute of Photonics and Electronics Academy of Science of the Czech Republic, v.v.i., 18251 Prague, Czech Republic;

[b]Laboratoire de Physique de la Matière Condensée, Université de Nice-Sophia-Antipolis, CNRS UMR 6622, Avenue Joseph Vallot, Parc Valrose, 06108 NICE CEDEX 2, France

[c]Now with Optoelectronics Research Centre, University of Southampton, SO17 1BJ, Southampton, United Kingdom

**ABSTRACT**

Silica-based thulium-doped fiber devices operating around 810 nm would extend the spectral range covered by high-power fiber devices. Using a comprehensive numerical model of the fiber we have shown that efficient lasing at 810 nm can be achieved for specific ranges of the laser cavity parameters in silica-based thulium-doped fibers with enhanced $^3H_4$ lifetime up to 58 µs as measured in our highly alumina-codoped fibers. We present optimization of the thulium-doped fiber and laser cavity parameters and also potential applications of the developed host material in amplifiers and broadband sources.

## 1. INTRODUCTION

Fiber lasers are one of the most spectacular achievements of contemporary optoelectronics and photonics. Variety of fiber lasers were demonstrated ranging from mighty continuous-wave (cw) systems with multi-kilowatt output powers to elegant femtosecond-pulse or tunable-narrow-line sources. Recently, high power fiber lasers have attained growing interest as they may potentially replace conventional "bulk" solid-state lasers in many applications [1]. The geometry of fiber lasers provides several advantages over the conventional solid-state lasers such as inherently excellent output beam quality and thermal management. Construction of fiber lasers may profit also from the fact that a variety of fiber optic components are readily available as they were developed for optical communications. Since kilometers of optical fiber can be drawn from a single preform, fiber lasers have potential of cheap mass production. With the cladding-pump technique, fiber lasers and amplifiers have proven to be remarkably efficient at converting the high-power, low-brightness output of diode lasers into high-power, high-brightness laser beams. The most important multi-watt fiber lasers were reported using double-clad (DC) fibers doped with ytterbium (1030-1100 nm), erbium and ytterbium (1540-1600 nm) and thulium (1860-2090 nm) [2]. In the field of thulium-doped fiber lasers (TDFLs), significant progress has been achieved in optimization of cross-relaxation process in the TDF in which one 790 nm pump photon can initiate generation of two lasing photons at ~2 µm. Power conversion efficiency of 65 % was reported using such a two-for-one process [3].

The amplification and lasing at 800 nm band has been already investigated using fluoride-based TDFs and output power of up to 2 W and 37% slope efficiency was achieved [4]. The output power was limited by the pump damage threshold of the fluoride fiber. The low-phonon-energy, fluoride-based fibers are known to provide high quantum efficiency of thulium laser transitions in contrast to high phonon energy silica fibers. However, usage of fluoride fibers results in difficulties with fabrication, hygroscopicity and aging of the host material and low pump power damage threshold. Furthermore, the fusion splicing to conventional silica-based fibers is impossible. This necessitates the use of mechanical butt-splices that are comparatively lossy and prone to damage under high power. High price of fluoride fibers is also an issue for many applications. These limitations are the main cause for which devices based on fluoride fibers are relatively rarely seen in practical applications despite numerous laboratory demonstrations, e.g., lasing with fluoride TDFs was reported at wavelengths around 455, 480, 810, 1470, 1900 and 2250 nm [5]. Thulium wide emission at ~1470 nm can be used for amplification in the telecommunication S-band (1460-1530 nm). It was primarily the application of thulium-doped fiber amplifiers (TDFAs) in S-band that recently spurred the development of alternative host materials with reduced phonon energy that possess long-term reliability. TDFAs based on soft tellurite [6] and antimony-silicate [7] glasses were demonstrated. Later, ways of increasing the TDFA gain and efficiency were investigated by modifying the environment of thulium ions only locally, e.g., by codoping the fiber core with oxides of bismuth [8], aluminium [9, 10] and gallium or germanium [11, 12].

---

[*] peterka@ufe.cz; phone +420 26 773 527; fax +420 284 680 222; www.ufe.cz

Recently, we have analyzed a compact thulium-doped silica-based fiber laser at around 810 nm, especially the effect of host materials (standard silica, highly alumina doped silica and ZBLAN) on the laser performance [13]. Single- and dual-wavelength pumping scheme were also discussed in terms of mitigating the possible photodarkening that might happen and inhibit the lasing action under upconversion pumping around 1060 nm. Although lasing action at 803 nm was observed in silica based fibers studied for amplification in the telecommunication S-band [14], the details about the actual output power characteristics were not reported. To our knowledge, silica-based thulium-doped fibers (TDFs) have not been used in high-power fiber lasers at ~ 800 nm and neither have they been theoretically analyzed before the Ref. 13. In this contribution we present detailed optimization of the thulium-doped fiber and laser cavity parameters and also potential applications of the developed host material in amplifiers and broadband sources.

The silica-based fiber lasers around 800 nm would extend the spectral range covered by high-power fiber lasers. The single-transversal mode, high-power laser source in the 800 nm spectral band is of interest for a variety of applications. The laser can be used for fiber sensors, instrument testing and for pumping of special types of lasers and amplifiers, e.g. the bismuth-doped lasers and TDFAs. Bismuth-doped fibers pumped around 800 nm may shift their gain to 1300 nm telecommunication band [15, 16], where highly reliable silica-based fiber amplifiers are still unavailable. An efficient fiber laser in 800 nm spectral region could potentially be used as a replacement for titanium sapphire laser in some applications. A high-power amplifier in the 800 nm band would be useful in the optical fiber communications and short-haul free-space communications. Although the laser diodes at this wavelength have been available for a long time, to our knowledge, commercially available single-mode laser diodes are limited to about 200 mW of output power in diffraction limited beam. It should not be confused with, e.g., laser diode stacks of ~kW output power, that are highly multimode with very high $M^2$ >1000 factor.

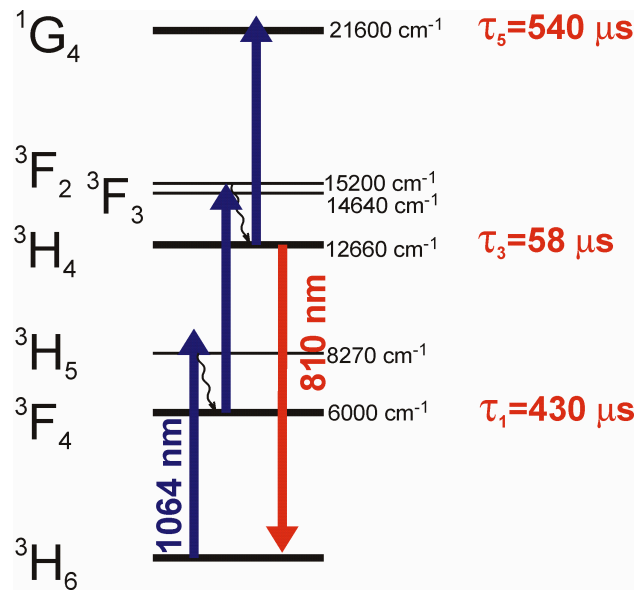

Figure 1. Thulium energy levels and the pumping scheme. Measured fluorescence lifetimes $\tau$ are shown right to the respective level involved in the model.

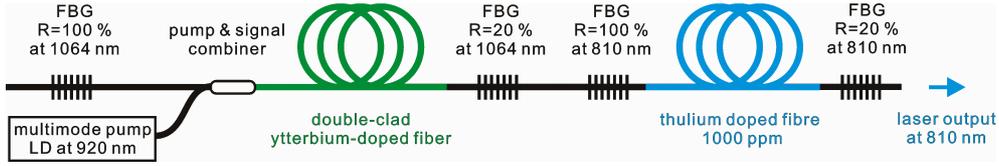

Figure 2. Setup of the fiber laser.

## 2. FIBER LASER SETUP AND PUMPING SCHEME

The proposed laser utilizes upconversion pumping scheme shown in Fig. 1 and has a compact all-fiber setup, example of which is shown in Fig. 2. Pumping at single wavelength in the range 1020-1090 nm would lead to second step of excited state absorption (ESA) to the $^1G_4$ level. This ESA is useless in our case. The upconversion pumping at single wavelength in the mentioned spectral range is studied as first choice because the high-power ytterbium-doped fiber lasers are readily available at these wavelengths. For pump wavelengths at the short-wavelength edge of the ytterbium emission spectrum one can use long-period fiber gratings in ytterbium-doped fibers to enhance power conversion efficiency of the pump laser as we have proposed recently [17, 18]. For experimental optimization of the pump wavelength within the range 1020-1090 nm the pump and signal combiner [19] can be used to facilitate changing the fiber Bragg gratings (FBG) that define the pump laser wavelength. The laser wavelength of the TDFL is selected by the two FBGs that have the wavelength of reflection at 810 nm. Overall cavity losses caused, e.g., by splice losses and insertion losses of FBGs, are estimated in the model to be 1 dB.

## 3. THEORETICAL MODEL AND FIBER CHARACTERISTICS

For theoretical investigation of thulium doped fiber device we used a comprehensive theoretical model of the TDF laser based on the fiber propagation and the laser rate equations. The model is spatially and spectrally resolved and it allows optimization parameters of the fiber laser, i.e., the fiber waveguide parameters, doping concentrations, fiber length, pump wavelength and power and laser cavity. The parameters of the TDF that are subject of optimization are schematically depicted in Fig. 3. The numerical model is based on solution of two sets of equations: differential equations describing the pump, signal and amplified spontaneous emission (ASE) that propagate in the fiber, so called propagation equations, and the set of rate equations describing the population of thulium ions in respective energy levels. Under steady state conditions the rate equations become a set of algebraic equations. The propagation and rate equations are solved simultaneously along the fiber. Since boundary conditions for the backward ASE are not known at the beginning of the fiber, an iterative process must be applied. The numerical model is described in details elsewhere [20, 21].

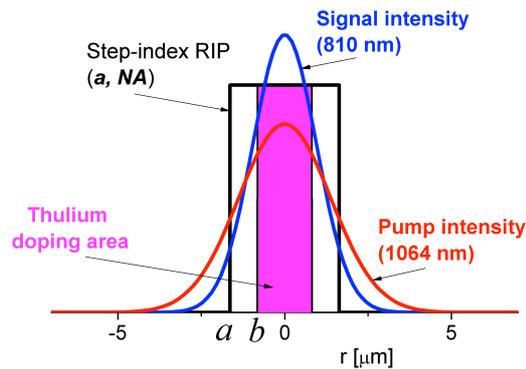

Figure 3. Refractive index profile (RIP) of the TDF with schematic representation of parameters that are subject of optimization: core radius *a*, numerical aperture *NA* and thulium doping radius *b*.

The fiber preform was fabricated by using the MCVD (Modified Chemical Vapor Deposition) and solution doping methods. From fiber attenuation we determined that the concentration of thulium was 40 ppm mol. The concentrations of $GeO_2$ and alumina were measured by electron microprobe analysis to be up to 4 and 10 mol %, respectively. The background loss of 0.1 dB/m was measured in spectral region apart from thulium absorption bands at around 1300 nm. The background loss is attributed mainly to relatively high concentration of OH groups of about 7.5 ppm. The fluorescence lifetimes measured in the TDF are shown in Fig. 1 next to the respective energy level. The $^3H_4$ lifetime of 58 µs in a highly alumina-codoped fiber represents a significant improvement compared to pure or weakly modified silica glass, where this lifetime is about 14 µs [22]. The fluorescence lifetime of the $^3H_4$ level was determined from the measured fluorescence decay. Since the decay curve was not possible to fit with a single exponential, the lifetime was determined as the time interval in which the fluorescence dropped to the 1/e of its peak value. Detailed description of the lifetime measurement and discussion were presented in [9, 10]. For the numerical modeling, we consider thulium concentration of 1000 ppm mol. With such a concentration, the fiber length can be relatively short while the cooperative quenching of thulium ions excited to the $^3H_4$ level can still be neglected [22]. The measured absorption and emission spectra around 800 nm in the fiber sample can be found in [10, 13]. The other spectroscopic parameters are taken from [20, 21].

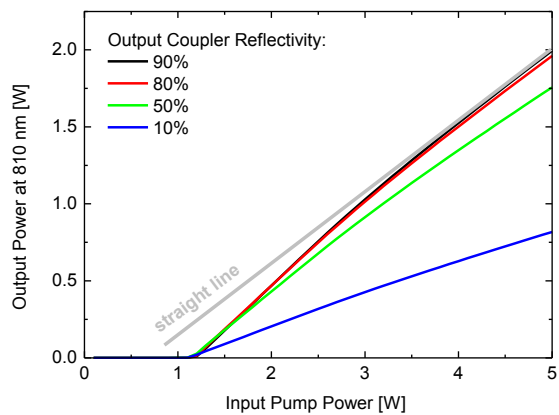

Figure 4. Laser output power characteristics vs. input pump power at 1064 nm for various output coupler ratios. The TDF parameters were as follows: length 4 m, background loss 0.1 dB/m, core radius is 1.7 µm and NA 0.2. The laser characteristics are not linear due to gradual population of the upper lying level $^1G_4$.

## 4. RESULTS AND DISCUSSION

By means of the numerical model we have optimized the laser cavity shown in Fig. 2 and the TDF parameters. The laser performance was optimized in terms of maximum output power at the optimum fiber length for a given input pump power. It would be more general to investigate the performance in terms of the laser threshold and slope efficiency. However, in the case of the upconversion pumping at around 1060 nm the second ESA to $^1G_4$ level leads to laser output rollover with higher pumping. At higher pump, more and more ions are promoted to the $^1G_4$ state and these ions do not contribute to the desired lasing transition. Example of the laser output characteristics can be seen in Fig. 4. The TDF parameters in this example were as follows: fiber length 4 m, background loss 0.1 dB/m, core radius 1.7 µm and numerical aperture 0.2. The whole core area was homogenously doped with thulium. Since the laser characteristics are not linear, we can not define unambiguously the slope efficiency and we investigate directly the laser output power. It can be inferred from the Fig. 4 that the output power varies only little for the output FBG reflectivity between 70-90%. Therefore we chose 80% reflectivity of the output FBG in further modeling of the laser. Firstly, we have analyzed the effect of fiber core radius $a$, numerical aperture $NA$ and then also the radius of thulium doping $b$. In Fig. 5, the laser output and the corresponding optimum length of the TDF are plotted for a range of fiber core radii and numerical apertures. Three individual graphs stand for three different background losses: (a) the ideal case of zero background losses, (b) the optimistic case of achievable background loss of 0.02 dB/m and (c) the last graph with background loss of 0.1 dB/m, measured in one of the developed fiber samples. The range of relevant core radii is limited to the interval of single-mode regime at the laser wavelength, i.e., where the expected effective cutoff wavelength is smaller than 800 nm. The output power is increasing with increasing NA, but the enhancement become less significant above NA=0.2. Since fabrication of TDF with higher NA may pose fabrication difficulties, we restrict further studies to fibers with NA=0.2 as a compromise. The corresponding optimal core radius is about 1.7 μm.

We have investigated also the influence of the thulium doping radius. In Fig. 6, the laser output and the corresponding optimum length of the TDF is plotted for a range of normalized doping radii $B=b/a$ and numerical apertures. The core radius $a$=1.7 μm was used in the calculations. Same as in Fig. 5, we investigated three values of background losses: 0, 0.02 and 0.1 dB/m. With small background losses that would allow for long TDF, the output power is slightly increasing for smaller doping area than the fiber core. However, for realistic background losses between 0.02 – 0.1 dB/m, the enhancement due to better overlap of the pump and laser signal is lost and the thulium doping in the whole fiber core allows for shorter TDF and consequently higher output powers. Therefore, we assume thulium doping across the whole area of the fiber core in the following calculations.

The laser output power vs. pump wavelength is shown in Fig. 7. In comparison with the TDFA for the telecommunications S-band (1460-1530 nm) with optimal pump wavelength around 1020 nm [23] the region of optimal pump wavelengths is shifted towards longer wavelengths because in this case high inversion between $^3H_4$ and $^3H_6$ levels and high pump absorption from $^3H_6$ level is desirable. Although the laser wavelength of 1064 nm is intended to be used in first experiments thanks to its availability, it can be seen from Fig. 7 that the 1064 nm is not the most efficient pump wavelength. The range of optimum wavelengths is between 1070-1120 nm. However, it should be noted that the key parameters for the pump power optimization are the pump ground- and excited-state absorption cross sections that were not measured for the specific fiber sample. Instead, the cross section spectra were taken from literature [21, 22]. Experimental optimization of the optimum pump wavelength would be highly desirable and that is why we propose the laser setup with easily interchangeable FBGs of the pump laser shown in Fig. 2. In order to show potential of the developed TDF core composition in fiber amplifier and broadband source we have studied theoretically performance of these devices. The parameters used in the modeling were the same as in optimized TDF for the fiber laser at 810 nm mentioned above. It should be noted that for both the TDFA and wideband ASE source the optimum parameters might be different but this optimization task is beyond the scope of this contribution. In contrast to the laser setup in Fig. 2, the amplifier uses wavelength division multiplexers to combine the signal and the pump and isolators to prevent laser oscillations. The results of the gain and noise figure calculations are summarized in Fig. 8 for signal input level of 100 μW. In the case of broadband source, the ASE spectra were calculated when no input signal is present, see Fig. 9. Similarly to the laser study described in [13], the fiber length is a very important parameter, especially in silica-based TDFs, because for longer TDF the radiation around 800 nm is gradually reabsorbed and the transition $^3F_4 \rightarrow {}^3H_6$ at around 2 μm becomes dominant.

## 5. CONCLUSIONS

Using a comprehensive numerical model of the TDF we have optimized the parameters of thulium-doped silica based fiber laser we have recently proposed. We have investigated the fiber length, core radius, numerical aperture, radius of the thulium-doped area, output coupler ratio and pump wavelength. We have shown that efficient lasing at 810 nm can be achieved using silica based Tm-doped fiber with enhanced $^3H_4$ lifetime for specific ranges of the fiber and laser cavity parameters. We have also investigated performance of the developed host material in the fiber amplifier and broadband ASE source. More than the 30 dB of gain of the amplifier and more than 500 mW output power of the ASE source with FWHM of 5 nm at 3 W of pump power were predicted.

This work was partially supported by CNRS Office of International Relations project No. PICS 5304 and by the Ministry of Education, Youth and Sports of the Czech Republic, project No. ME10119 "FILA". LPMC is with GIS 'GRIFON' (CNRS, www.unice.fr/GIS ).

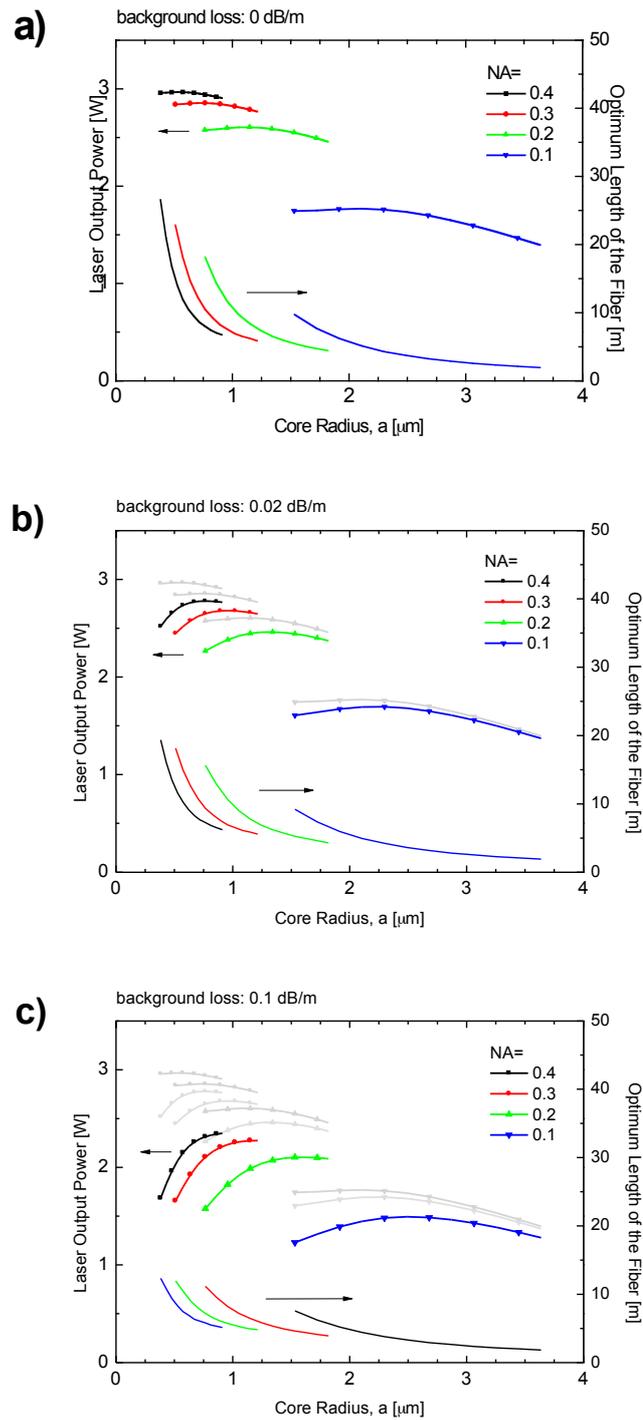

Figure 5. Laser output power and optimum fiber length vs. core radius for several values of the NA and background loss. Only the relevant core radii where the effective cutoff wavelength of the fiber shall be below 800 nm were taken into account. The doping radius was set to $b=1$. The light grey curves are the laser output power results from the preceding figure(s), for easy comparison between the three graphs.

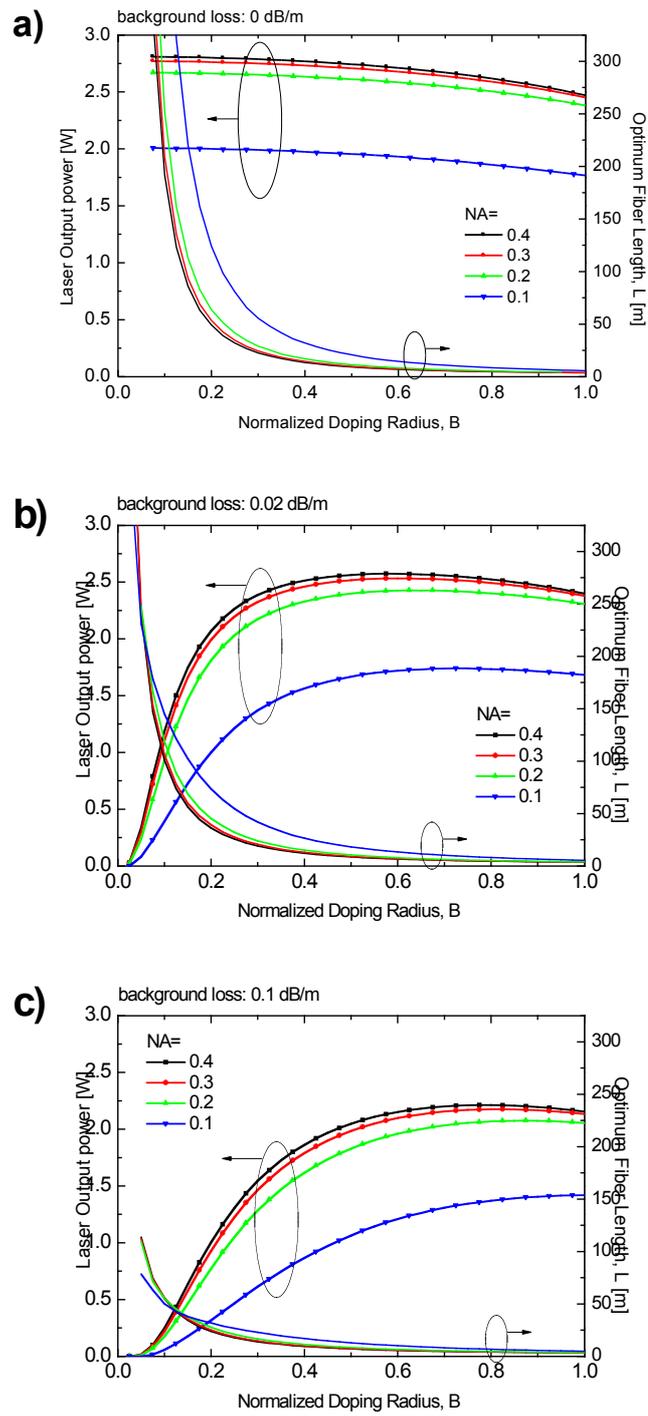

Figure 6. The effect of thulium doping area radius *b* on the laser performance for three values of background loss. The core radius was kept *a*=1.7 μm.

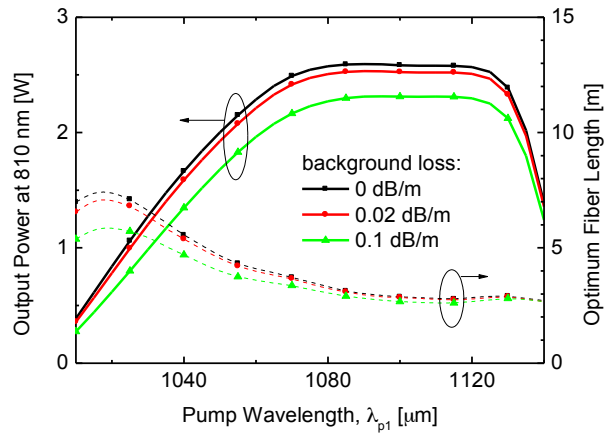

Figure 7. Optimization of the pump wavelength for several values of background loss and pump power of 5 W.

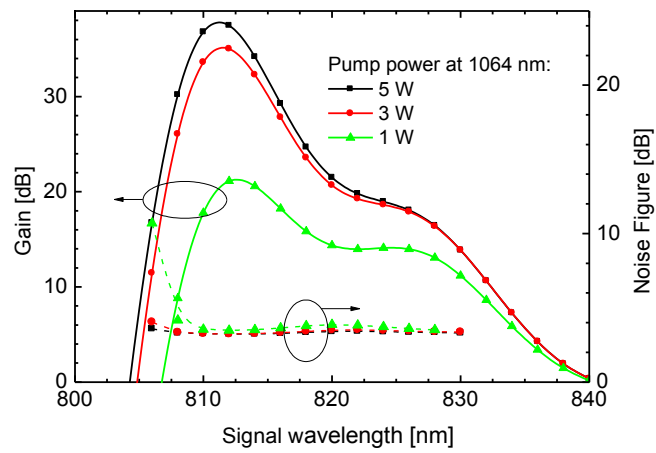

Figure 8. Gain and noise figure of the TDFA for several values of pump power at 1064 nm. Signal input power is 100 μW.

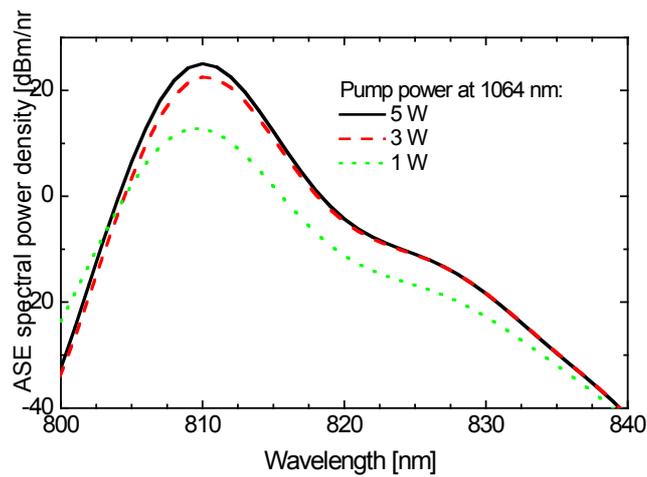

Figure 9. Backward ASE spectra for several values of pump power at 1064 nm and TDF length of 4 m.

# REFERENCES

[1] Tünnermann A., Schreiber T., and Limpert J., "Fiber lasers and amplifiers: an ultrafast performance evolution," Appl. Opt. 49, F71-F78 (2010).
[2] Nilsson J., Clarkson W. A., Selvas R., Sahu J. K., Turner P. W., Alam S. U. and Grudinin A. B., "High-power wavelength-tunable cladding-pumped rare-earth-doped silica fiber lasers", Optical Fiber Technology 10, 5-30 (2004).
[3] Moulton P. F., Rines G. A., Slobodtchikov E. V., Wall K. F., Frith G., Samson B. and Carter A. L. G., "Tm-doped fiber lasers: Fundamentals and power scaling", IEEE J. Sel. Topics in Quantum Electronics 15, 85-92 (2009).
[4] Dennis M. L., Dixon J. W. and Aggarwal I., "High power upconversion lasing at 810 nm in Tm:ZBLAN fibre", El. Lett. 30, 136-137 (1994).
[5] Digonnet M.J.F., [Rare earth doped fiber lasers and amplifiers], New York, M. Dekker, (1993).
[6] Taylor E. R. M., Ng L. N., Nilsson J., Caponi R., Pagano A., Potenza M. and Sordo B., "Thulium-doped tellurite fiber amplifier", IEEE Photonics Technol. Lett. 16, 777-779 (2004).
[7] Samson B. N., Traynor N. J., Walton D. T., Ellison A. J. G., Minelly J. D., Trentelman J. P. and Dickinson J. E., "Thulium doped silicate fiber amplifier at 1460-1520 nm", Optical Amplifiers and Their Applications, July 9-12, Quebec, Canada, 44, 247-249 (2000).
[8] Ohara S., Sugimoto N., Kondo Y., Ochiai K., Kuroiwa Y., Fukasawa Y., Hirose T., Hayashi H. and Tanabe S., "$Bi_2O_3$-based glass for S-band amplification", Proc. SPIE 4645, 8-15 (2002).
[9] Blanc W., Sebastian T.L., Dussardier B., Michel C., Faure B., Ude M. and Monnom G., "Thulium environment in a silica doped optical fibre", J. Non-Cryst. Solids 354, 435-439 (2008).
[10] Faure B., Blanc W., Dussardier B. and Monnom G., "Improvement of the $Tm^{3+}:^3H_4$ level lifetime in silica optical fibers by lowering the local phonon energy", J. Non-Cryst. Solids 353, 2767-2773 (2007).
[11] Cole B. and Dennis M. L., "S-band amplification in a thulium doped silicate fiber",OFC'01, Anaheim, USA, TuQ3 (2001).
[12] Watekar P. R., Ju S., and Han W. -T., "A Novel 800 nm Double-Clad Optical Fiber Amplifier with 980 nm Pumping," Proc. of Opt. Amplifiers and Their Applications, p. OMC5 (2006).
[13] Peterka P., Kasik I., Dhar A., Dussardier B. and Blanc W., "Theoretical analysis of fiber lasers emitting around 810 nm based on thulium-doped silica fibers with enhanced $^3H_4$ level lifetime", In Europhysics Conference Abstracts 34C, 4$^{th}$ EPS-QEOD Europhoton conference, Germany, WeP5, (2010).
[14] Dennis M. L. and Cole B., "Amplification device utilizing thulium doped modified silicate optical fiber", US patent No. 6924928, August 2, (2005).
[15] Dianov E.M., Firstov S.V., Khopin V.F., Guryanov A.N., Bufetov I.A., "Bi-doped fibre lasers and amplifiers emitting in a spectral region of 1.3 μm", Quantum Electronics 38, 615-617 (2008).
[16] Bufetov I. and Dianov E. M., "Bi-doped fiber lasers", Laser Physics Letters 6, 487–504 (2009).
[17] Peterka P., Maria J., Dussardier B., Slavik R., Honzatko P., and Kubecek V., "Long-period fiber grating as wavelength selective element in double-clad Yb-doped fiber-ring lasers", Laser Physics Letters 6, 732-736 (2009).
[18] Peterka P. and Slavik R., "Extension of the double-clad Yb-doped fiber laser oscillation range thanks to long-period fiber grating filters", CLEO-Europe, Munchen, Germany, p. CJ.P.11-THU (2009).
[19] Peterka P., Kubecek V., Dvoracek P., Kasik I. and Matejec V., "Experimental demonstration of novel end-pumping method for double-clad fiber devices", Opt. Lett. 31, 3240-3242 (2006).
[20] Peterka P., Blanc W., Dussardier B., Monnom G., Simpson D. A. and Baxter G. W., "Estimation of energy transfer parameters in thulium- and ytterbium-doped silica fibers", In Proc. SPIE 7138, p. 7138K (2008).
[21] Peterka P., Faure B., Blanc W., Karasek M. and Dussardier B., "Theoretical modelling of S-band thulium-doped silica fibre amplifiers", Optical and Quantum Electronics 36, 201-212 (2004).
[22] Jackson S. D. and King T. A., "Theoretical modeling of Tm-doped silica fiber lasers", J. Lightwave Technol. 17, 948-956 (1999).
[23] Peterka P., Kasik I., Matejec V., Blanc W., Faure B., Dussardier B., Monnom G. and Kubecek V., "Thulium-doped silica-based optical fibers for cladding-pumped fiber amplifiers", Optical Materials 30, 174-176 (2007).